\documentclass[twocolumn]{aastex61}

\usepackage{url}
\usepackage{ulem}
\usepackage{CJKutf8}

\newcommand{\appropto}{\mathrel{\vcenter{\offinterlineskip\halign{\hfil$##$\cr\propto\cr\noalign{\kern2pt}\sim\cr\noalign{\kern-2pt}}}}}
\newcommand{\bjdtdb}{\ensuremath{\rm {BJD_{TDB}}}}
\newcommand{\feh}{\ensuremath{\left[{\rm Fe}/{\rm H}\right]}}
\newcommand{\teff}{\ensuremath{T_{\rm eff}}}

\newcommand{\msun}{\ensuremath{\,M_\Sun}}
\newcommand{\rsun}{\ensuremath{\,R_\Sun}}
\newcommand{\lsun}{\ensuremath{\,L_\Sun}}
\newcommand{\mj}{\ensuremath{\,M_{\rm J}}}
\newcommand{\rj}{\ensuremath{\,R_{\rm J}}}
\newcommand{\fave}{\langle F \rangle}
\newcommand{\fluxcgs}{10$^9$ erg s$^{-1}$ cm$^{-2}$}

\begin{document}


\title{Transiting Exoplanet Monitoring Project (TEMP). I. Refined System Parameters and Transit Timing Variations of HAT-P-29b}  

\correspondingauthor{Songhu Wang}
\email{song-hu.wang@yale.edu}

\author{Songhu Wang}
\affiliation{Department of Astronomy, Yale University, New Haven, CT 06511, USA}
\affiliation{\textit{51 Pegasi b} Fellow}

\author{Xian-Yu Wang}
\affiliation{Key Laboratory of Optical Astronomy, National Astronomical Observatories, Chinese Academy of Sciences, Beijing 100012, China}
\affiliation{University of Chinese Academy of Sciences, Beijing, 100049, China}

\author{Yong-Hao Wang}
\affiliation{Key Laboratory of Optical Astronomy, National Astronomical Observatories, Chinese Academy of Sciences, Beijing 100012, China}
\affiliation{University of Chinese Academy of Sciences, Beijing, 100049, China}

\author{Hui-Gen Liu}
\affiliation{School of Astronomy and Space Science and Key Laboratory of Modern Astronomy and Astrophysics in Ministry of Education, Nanjing University, Nanjing 210093, China}

\author{Tobias C. Hinse}
\affiliation{Korea Astronomy \& Space Science Institute, 305-348 Daejeon, Republic of Korea}

\author{Jason Eastman}
\affiliation{Harvard-Smithsonian Center for Astrophysics, Garden Street, Cambridge, MA 02138 USA}

\author{Daniel Bayliss}
\affiliation{Department of Physics, University of Warwick, Gibbet Hill Road, Coventry CV4 7AL, UK}

\author{Yasunori Hori}
\affiliation{National Astronomical Observatory of Japan, NINS, 2-21-1 Osawa, Mitaka, Tokyo 181-8588, Japan}
\affiliation{Astrobiology Center, 2-21-1 Osawa, Mitaka, Tokyo, 181-8588, Japan}

\author{Shao-Ming Hu}
\affiliation{Shandong Provincial Key Laboratory of Optical Astronomy and Solar-Terrestrial Environment, Institute of Space Sciences, Shandong University, Weihai 264209, China}

\author{Kai Li}
\affiliation{Shandong Provincial Key Laboratory of Optical Astronomy and Solar-Terrestrial Environment, Institute of Space Sciences, Shandong University, Weihai 264209, China}

\author{Jinzhong Liu}
\affiliation{Xinjiang Astronomical Observatory, Chinese Academy of Sciences, Urumqi, Xinjiang 830011, China}

\author{Norio Narita}
\affiliation{National Astronomical Observatory of Japan, NINS, 2-21-1 Osawa, Mitaka, Tokyo 181-8588, Japan}
\affiliation{Astrobiology Center, 2-21-1 Osawa, Mitaka, Tokyo, 181-8588, Japan}
\affiliation{Department of Astronomy, The University of Tokyo, 7-3-1 Hongo, Bunkyo-ku, Tokyo 113-0033, Japan}

\author{Xiyan Peng}
\affiliation{Key Laboratory of Optical Astronomy, National Astronomical Observatories, Chinese Academy of Sciences, Beijing 100012, China}

\author{R. A. Wittenmyer}
\affiliation{University of Southern Queensland, Computational Engineering and Science Research Centre, Toowoomba, Queensland 4350, Australia}

\author{Zhen-Yu Wu}
\affiliation{Key Laboratory of Optical Astronomy, National Astronomical Observatories, Chinese Academy of Sciences, Beijing 100012, China}
\affiliation{University of Chinese Academy of Sciences, Beijing, 100049, China}

\author{Hui Zhang}
\affiliation{School of Astronomy and Space Science and Key Laboratory of Modern Astronomy and Astrophysics in Ministry of Education, Nanjing University, Nanjing 210093, China}

\author{Xiaojia Zhang}
\affiliation{Department of Earth Sciences, The University of Hong Kong, Pokfulam Road, Hong Kong }

\author{Haibin Zhao}
\affiliation{Key Laboratory of Planetary Sciences, Purple Mountain Observatory, Chinese Academy of Sciences, Nanjing 210008, China}

\author{Ji-Lin Zhou}
\affiliation{School of Astronomy and Space Science and Key Laboratory of Modern Astronomy and Astrophysics in Ministry of Education, Nanjing University, Nanjing 210093, China}

\author{George Zhou}
\affiliation{Harvard-Smithsonian Center for Astrophysics, Garden Street, Cambridge, MA 02138 USA}

\author{Xu Zhou}
\affiliation{Key Laboratory of Optical Astronomy, National Astronomical Observatories, Chinese Academy of Sciences, Beijing 100012, China}

\author{Gregory Laughlin}
\affiliation{Department of Astronomy, Yale University, New Haven, CT 06511, USA}

\begin{abstract} 
 We report the photometry of six transits of the hot Jupiter HAT-P-29b obtained from 2013 October to 2015 January. We analyze the new light curves, in combination with the published photometric, and Doppler velocimetric, and spectroscopic measurements, finding an updated orbital ephemeris for the HAT-P-29 system, $T_{\rm C}[0]= 2456170.5494(15)\,[\rm{BJD_{TDB}}]$ and $P=5.723390(13) \,{\rm days}$. It is $17.63\,{\rm s}$ ($4.0\,\sigma$) longer than the previously published value, amounting to errors exceeding $2.5\,\rm{hrs}$ at the time of writing (on UTC 2018 June 1). The measured transit mid-times for HAT-P-29b show no compelling evidence of timing anomalies from a linear model, which rules out the presence of a perturbers with masses greater than $0.6$, $0.7$, $0.5$, and $0.4\,{\rm M_\oplus}$ near the $1:2$, $2:3$, $3:2$, and $2:1$ resonances with HAT-P-29b, respectively. 
\end{abstract}


\section{Introduction}

High-precision photometric follow-up observations permit the refined determination of physical properties (especially the radii) of known transiting exoplanets (e.g., \citealt{Holman2006, Southworth2009, WangY2017, WangX2018}). An accumulation  of these data provide new insights into the distribution of planetary interior structures, formation and evolution processes.

Follow-up observations are also required to update and maintain planetary orbital ephemerides \citep{Wang2018a}, which are 
needed in order to confidently schedule in-transit follow-up observations (e.g., Rossiter-Mclaughlin effect measurements: \citealt{Wang2018b}, or atmospheric transmission spectra observations: \citealt{knutson2012}). 

\begin{figure*}
\vspace{0cm}\hspace{0cm}
\includegraphics[width=\textwidth]{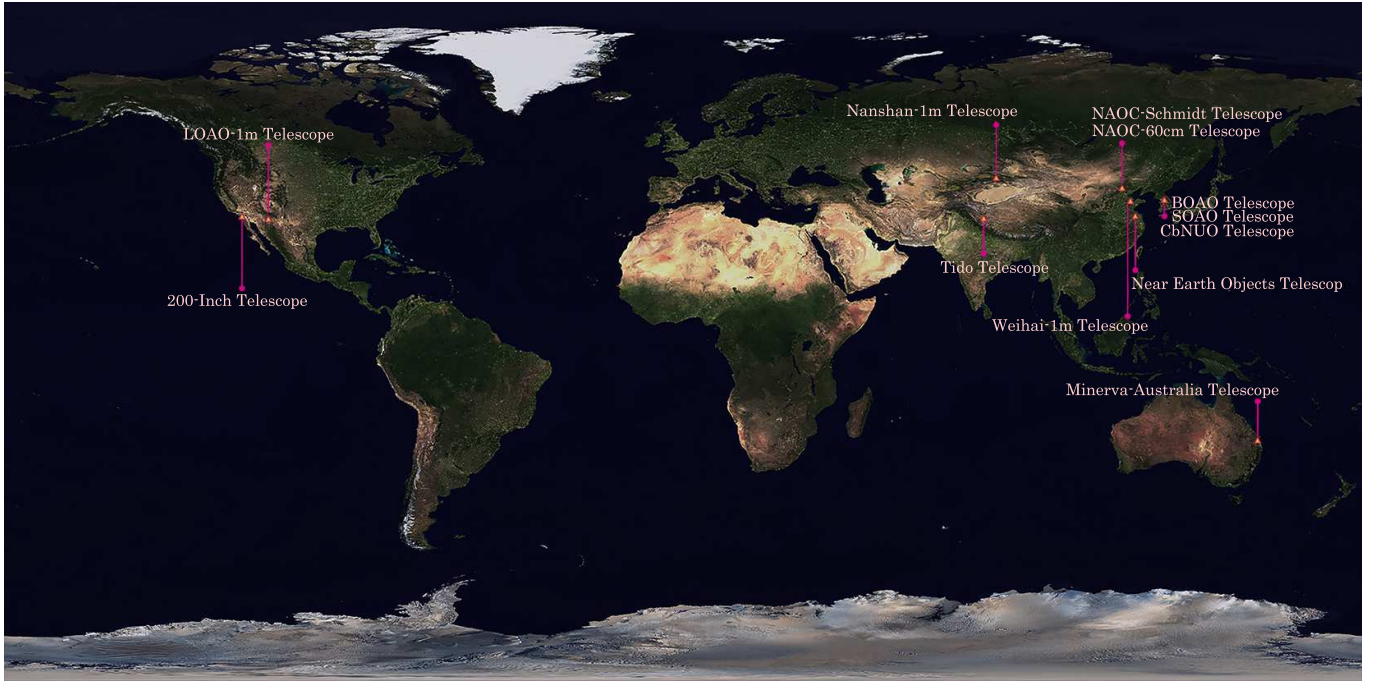}
\caption{TEMP network locations. The map shows the latitude and longitude coverage of the TEMP observatories. 
\label{fig1}}
\end{figure*}

Moreover, high-precision photometric follow-up observations, and by extension, accurate measurements of transit timing variations (TTVs) of known hot Jupiters, offer a powerful tool for the detection of hot Jupiter companion planets with masses comparable to Earth's \citep{miralda2002, agol2005, holman2005}. It will provide a key zeroth-order test of the competing mechanisms of hot Jupiter formation \citep{Millholland2016}.

Photometric follow-up observations are also often needed before definitive TTV-determined masses can be achieved for K2 (and in the near future, TESS) planets \citep{Grimm2018, Wang2017}. The $K2$ \citep{Howell2014} and upcoming $TESS$ \citep{Ricker2015} missions only monitor each target field for $\sim80$ and $\sim27$ days, respectively. The timescales associated with TTV signals, however, are typically several years \citep{Agol2017, Wu2018}. 

Finally, high-precision multi-band transit photometry is also a powerful diagnostic tool for exploring the atmospheric properties of close-in planets, notably, the atmospheric compositions and meteorological conditions associated with clouds and hazes (e.g., \citealt{Sing2016}). 

\setlength{\tabcolsep}{1.1pt}
\begin{deluxetable*}{llccccc}
\tabletypesize{\scriptsize}
\tablewidth{0pt}
\tablecaption{\label{table1} TEMP Network Telescopes}
\tablehead{\colhead{Telescope}  &  \colhead{Observatory}   &  \colhead{Longitude} & \colhead{Latitude} &  \colhead{Aperture}  &  \colhead{FOV}  &  \colhead{Pixel Scale}  \\
}
\startdata
200-Inch Telescope/WIRC &  Palomar Observatory & $116^{\circ}51'54''$W & $33^{\circ}21'21''$N &  5m & $8.7'\times8.7'$ & 0.25 \\  
NAOC-Schmidt Telescope &  National Observatory of China & $117^{\circ}34'30''$E & $40^{\circ}23'39''$N & 0.6/0.9m & $94'\times94'$ & 1.37''/pixel \\  
NAOC-60cm Telescope &  National Observatory of China & $117^{\circ}34'30''$E & $40^{\circ}23'39''$N & 0.6m & $17'\times17'$ & 1.95''/pixel \\  
Weihai-1m Telescope &  Shandong University/Weihai Observatory & $122^{\circ}02'58''$E & $37^{\circ}32'09''$N &  1.0m & $12'\times12'$ & 0.35''/pixel \\  
Tido Telescope  &  Nanjing University/Ali Observatory & $80^{\circ}05'57.14''$E &  $32^{\circ}29'46.26''$N  & 1,0m  & $3^{\circ} \times 3^{\circ}$ & 1.76''/pixel \\  
Nanshan-1m Telescope  &  Xinjiang Observatory & $87^{\circ}10'30''$E  & $43^{\circ}28'24.66''$N & 1.0m & $1.3^{\circ}\times1.3^{\circ}$  & 1.14''/pixel \\  
Near Earth Objects Telescope  &  Purple Mountain Observatory & $118^{\circ}28'$E & $32^{\circ}44'$N &  1.2m & $3.0^{\circ}\times3.0^{\circ}$ & 1.03''/pixel \\  
Minerva-Australia Telescope  &  USQ/Mt Kent Observatory & $151^{\circ}51'19.5''$E &  $270^{\circ}47'52.3''$E & 0.7m  & $21'\times21'$  & 0.6''/pixel  \\ 
BOAO Telescope & Bohyunsan Optical Astronomy Observatory & $128^{\circ} 58' 35''\,\rm E$ & $36^{\circ} 09' 53''\,\rm N$ & 1.8m & $14.6' \times 14.6'$ & 0.21''/pixel \\
SOAO Telescope & Sobaeksan Optical Astronomy Observatory & $128^{\circ} 27' 25''\,\rm E$ & $36^{\circ} 56' 13''\,\rm N$ & 0.6m & $17.6' \times 17.6'$ & 0.52''/pixel \\
CbNUO Telescope & Chungbuk National University Observatory & $127^{\circ} 28' 31''\,\rm E$ & $ 36^{\circ} 46' 53''\,\rm N$ & 0.6m & $72' \times 72'$ & 1.05''/pixel \\
LOAO-1m Telescope & Mt. Lemmon Optical Astronomy Observatory & $249^{\circ} 12' 41''\,\rm E$ & $32^{\circ} 26' 32''\,\rm N$ & 1.0 & $22.2' \times 22.2'$ & 0.64''/pixel  \\
\enddata
\end{deluxetable*}

For these reasons, we have initiated the Transiting Exoplanet Monitoring Project (TEMP) to gather long-term, high-quality photometry of exoplanetary transits with 1-meter-class ground-based telescopes (see Figure~\ref{fig1} and Table~\ref{table1} for the TEMP network locations and telescopes). The scientific goals are:
 
 \begin{itemize}
\item Identify and characterize undetected planets interleaved among known transiting planets via TTVs.

\item Refine the orbital and physical parameters of the known transiting planets discovered with ground-based photometric surveys, which usually only received a handful of photometric follow-up observations.

\item Make definitive estimates of planetary masses in multi-transiting systems discovered with $K2$ and $TESS$ via TTVs.

\item Characterize exoplanetary  compositions and atmospheric properties with multi-band photometry.
\end{itemize}

\begin{figure}
\vspace{0cm}\hspace{0cm}
\includegraphics[width=\columnwidth]{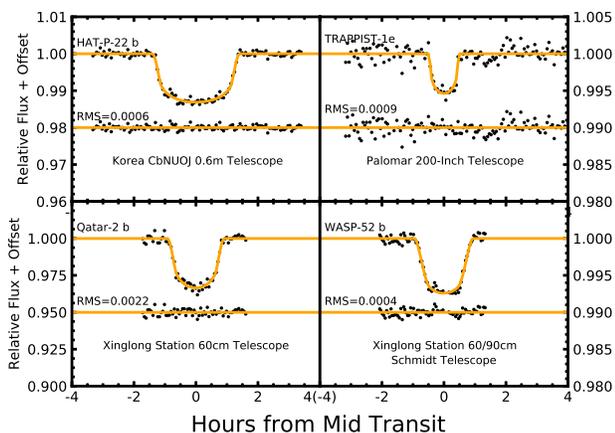}
\caption{Example light curves (Black points) from TEMP compared to the best-fitting models (Yellow lines).
The residuals are offset from zero to the base of each panel for clarity. 
\label{fig2}}
\end{figure}

To date, $\sim 300$ light curves of about $60$ transiting exoplanets have been obtained through the TEMP network (\citealt{WangY2018}, In prep.). The light curves (see Figure~\ref{fig2} for examples) have a typical photometric precision ranging from 1 to $2\,\rm{mmag}$, depending on weather and the stellar magnitude. In the best cases, sub-mmag photometric precision has been achieved. 

Here, we present one of our first scientific results, namely a refined characterization of the HAT-P-29 planetary system.

The transiting hot Jupiter HAT-P-29b was discovered by \citet{buchhave2011} under the auspices of the HATNet project. Although extended Doppler velocity monitoring shows evidence for the existence of a distant outer companion in the system \citep{knutson2014}, HAT-P-29b is otherwise a comparatively normal transiting exoplanetary system consisting of a $1.2\, {\rm M_{\odot}}$ star circled by a $0.78\,{\rm M_{JUP}}$ planet with an orbital period of $5.72\,{\rm days}$. The relatively long period introduces challenges for ground-based follow-up using meter-class telescopes. The photometric characterization of HAT-P-29b in the discovery work rested on only two partial follow-up transit light curves, and the discovery light curve which is of limited quality. Moreover, four additional Doppler velocimetric measurements (there are eight in discovery paper) were obtained by \citet{knutson2014} for this system. \citet{Torres2012} also improved the spectroscopic properties of the host star for this system.

Here, we report the first photometric transit follow-up of HAT-P-29b since the discovery work, covering six transits (only two of these are complete, however). This new material, coupled with all archival photometric \citep{buchhave2011}, spectroscopic \citep{Torres2012}, and Doppler velocimetric data \citep{buchhave2011, knutson2014}, permits refinement of the planetary orbital and physical properties. By analyzing the transit mid-times of all available follow-up light curves (six from this work, and two from \citealt{buchhave2011}), we effectively constrain the parameter space of the potential nearby perturbers.

We proceed in the following manner: 
In \S2, we describe the new photometric observations and their reduction.
\S3 details the technique we used to estimate the system parameters.
\S4 discusses our results and some implications.
A brief summary of this work is presented in \S5.

\section{Observation and Data Reduction}

\setlength{\tabcolsep}{1.1pt}
\begin{deluxetable*}{ccccccccc}
\tabletypesize{\scriptsize}
\tablewidth{0pt}
\tablecaption{Log of Observations}
\tablehead{
\colhead{Date}  &  \colhead{Time}   &  \colhead{Telescope} & \colhead{Filter} &  \colhead{Number of exposures}  &  \colhead{Exposure time}  &  \colhead{Airmass}  &  \colhead{Moon Phase} &  \colhead{Scatter}\tablenotemark{a}\\
\colhead{(UTC)}  &  \colhead{(UTC)}   &  \colhead{}          & \colhead{}       &  \colhead{}               &  \colhead{(second)}       &  \colhead{}         &  \colhead{}   &  \colhead{}  \\
}
\startdata
2013 Nov 14 & { }{ }13:34:55-19:32:22 & { }{ }NAOC-Schmidt &   $R$ &     190  &   80  &   1.02-1.52  & 0.91 & 0.0023 \\
2013 Dec 07 & { }{ }11:31:53-16:18:58 & { }{ }NAOC-Schmidt &   $R$ &     108  &   100 &   1.02-1.19  & 0.29 &0.0019 \\
2013 Dec 30 & { }{ }09:56:14-15:59:01 & { }{ }NAOC-Schmidt &   $R$ &     216  &   60  &   1.02-1.39   & 0.05 & 0.0019 \\
2014 Feb 08 & { }{ }10:49:06-13:51:03 & { }{ }NAOC-Schmidt &   $R$ &     109  &   60  &   1.07-1.50   & 0.68 &0.0026 \\
2014 Nov 21 & { }{ }14:00:18-18:55:51 & { }{ }NAOC-Schmidt &   $R$ &     332  &   35  &    1.02-1.46  & 0.01 & 0.0037 \\
2015 Jan  06 & { }{ }10:17:43-15:25:02 & { }{ }Weihai-$1\,{\rm m}$                           &   $V$ &     230   &   60  &    1.02-1.33  & 0.98 & 0.0023 \\
\enddata
\tablenotetext{a}{Scatter represents the RMS of the residuals from the best-fitting transit model. }
\label{table2}
\end{deluxetable*}

Five transits of HAT-P-29b, between 2013 October and 2014 November were observed in a Cousins $R$ filter with the $60/90\,{\rm cm}$ Schmidt telescope at Xinglong Station (${\rm 117^\circ 34'30''E}$, $\rm{40^\circ 23'39''N}$) of the National Astronomical Observatories of China (NAOC). 
The telescope is equipped with a $4{\rm K} \times 4{\rm K}$ CCD that gives a $94' \times 94'$ field of view (FOV). A $512 \times 512$ pixel (approximately $11.7' \times 11.7'$) subframe was used to reduce the readout time from 93 to $4\,{\rm s}$, significantly increasing the duty-cycle of the observations. For our observations, the images were not binned, giving a pixel scale of $1.38''\,{\rm pixel ^{-1}}$.
For full details of this telescope, we refer the reader to \citet{zhou1999, zhou2001}.

A sixth transit, obtained on UTC 2015 January 6, was observed through a Johnson $V$ filter using the 1-m telescope operated at Weihai Observatory (${\rm 122^\circ 02'58.6''E, 37^\circ 32'09.3''N}$) of Shandong University, China. The telescope has a $2{\rm K} \times 2{\rm K}$  CCD with a $12' \times 12'$ FOV. No windowing or binning was used, resulting in a pixel scale of $0.35''\,{\rm pixel ^{-1}}$, and a readout/reset time between exposures of $15\,\rm{s}$. For further instrumental details of this telescope, see \citet{Hu2014}. 

To maintain a high signal-to-noise ratio (${\rm SNR}>1000$) for both the target and comparison stars, exposure times were varied from 35 to $100\,{\rm s}$, depending on atmospheric conditions. Nevertheless, exposure times were not changed during the ingress or egress phases to avoid adversely affecting the transit timing. The mid-exposure time is recorded in the image header, and is synchronized with the USNO master clock time\footnote{$\,$http://tycho.usno.navy.mil/.} at the beginning of each night. The intrinsic error for all recorded times in the image headers is estimated to be less than $1\,$s. 
The recorded time stamps are converted from ${\rm JD_{UTC}}$ to ${\rm BJD_{TDB}}$ using the techniques of \citet{eastman2010}.
A summary of our observations is given in Table~\ref{table2}.

\begin{deluxetable}{cccc}
\tablewidth{0pt}
\tablecaption{Photometry of HAT-P-29}
\tablehead{
\colhead{${\rm BJD_{TDB}}$\tablenotemark{a}} & \colhead{Relative Flux} & \colhead{Uncertainty}  & \colhead{Filter} 
}
\startdata
      2456611.083603   & 0.9995 &    0.0023    & $R$ \\
      2456611.084783   & 0.9965 &    0.0023    & $R$ \\
      2456611.085976   & 0.9996 &    0.0023    & $R$\\
      2456611.087168   & 1.0002 &    0.0023    & $R$ \\
      2456611.088348   & 1.0004 &    0.0023   & $R$ \\
      2456611.089540   & 1.0019 &    0.0023   & $R$ \\ 
      2456611.090721   & 0.9997 &    0.0023   & $R$\\
      ...              & ...    &    ...      & ... \\
\enddata
\tablenotetext{a}{ The time stamps are based on the Barycentric Julian Date (BJD) in Barycentric Dynamical Time (TDB). 
The timings throughout the paper are placed on the ${\rm BJD_{TDB}}$ time system.}
\label{table3}
\end{deluxetable}

\begin{figure}
\vspace{0cm}\hspace{0cm}
\includegraphics[width=\columnwidth]{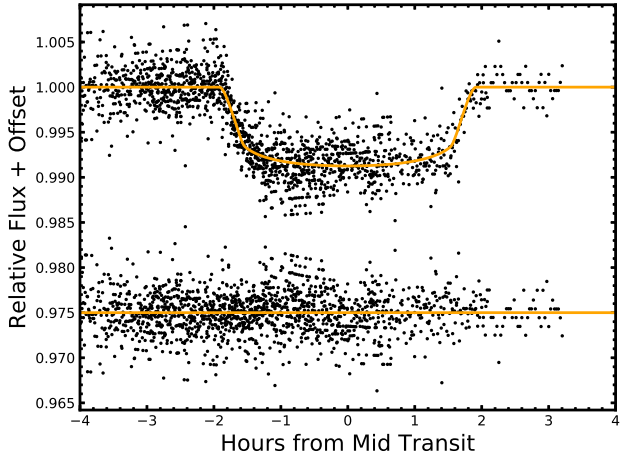}
\caption{Phased follow-up light curves of HAT-P-29 (two from \citealt{buchhave2011}, and six from this work).
These light curves are fitted simultaneously with the Doppler velocity measurements from \citet{buchhave2011} and \citet{knutson2014} to estimate the system parameters (see Figure~\ref{fig4} and \S3). The solid orange line shows the best-fitting model and the residuals of fit are plotted below. 
\label{fig3}}
\end{figure}

All data are bias-corrected and flat-fielded using standard routines.
Aperture photometry is then performed using SExtractor \citep{Bertin1996}.
The final differential light curves are obtained from weighted ensemble photometry. 
The aperture sizes and the choice of comparison stars are optimized to minimize the out-of-transit root-mean-square (RMS) scatter.
The resulting light curves are given in Table~\ref{table3}, and are compared in Figures~\ref{fig3} and \ref{fig5} to the best-fitting model\footnote{$\,$Table~\ref{table3} is available in its entirety on http://casdc.china-vo.org/archive/TEMP/HAT-P-29/.}. The scatter of the residuals from the best-fitting model in these light curves varies from 1.9 to $3.7\,$mmag.

\section{Light Curve Analysis}

To re-estimate the system parameters of HAT-P-29, we performed two global fittings, one in which we only used follow-up light curves (six from this work, and two from \citealt{buchhave2011}), the second included the HATNet discovery photometry. We simultaneously model photometric data, together with high-precision RV measurements  \citep{buchhave2011, knutson2014} acquired with Keck/HIRES using EXOFAST \citep{Eastman2013}.

EXOFAST fits each data set independently to determine an error scaling and preliminary best fit, then uses the Differential Evolution Markov Chain Monte Carlo (DE-MCMC; \citealt{Braak2006}) to refine the best global fit (including all data sets, the \citealt{Torres2008} relation to determine the stellar parameters, and a loose prior on the limb darkening from \citealt{Claret2011}) and characterizes the uncertainties. We allow the eccentricity to float, although the fitted eccentricity is consistent with zero. The chains are run until well mixed, as described in \citet{Eastman2013} and the results are summarized in Table~\ref{table4}.

We imposed priors on the all transit and RV parameters -- the orbital period ($P$) of $5.723186 \pm 0.000049\,{\rm days}$,  the planet-to-star radius ratio ($R_{\rm P}/R_{\ast}$) of $0.0927 \pm 0.0028$, the scaled semimajor axis ($a/R_{\ast}$) of $11.70_{-0.97}^{+0.71}$, the inclination ($i$) of $87.1^{\circ}\,_{-0.7^{\circ}}^{+0.5^{\circ}}$,  the eccentricity ($e$) of $0.095 \pm 0.047$, the argument of periastron ($\omega_*$) of $169^{\circ} \pm 30^{\circ}$, the RV semi-amplitude ($K$) of $78.3\,{\rm m\,s^{-1}} \pm 5.9 \,{\rm m\,s^{-1}}$, -- from discovery paper \citep{buchhave2011}. We also set priors on the stellar spectroscopic parameters -- the stellar effective temperature ($T_{\rm eff}$) of $6086\pm 69 {\,\rm K}$, the surface gravity (${\rm log}\,g$) of $4.34 \pm 0.06$, metallicity ($\rm{[Fe/H]}$) of $0.14 \pm 0.08$ -- from \citet{Torres2012}. Moreover, we adopted wavelength dependent limb-darkening coefficients 
{$\mu_{1,R}=0.323$, and $\mu_{2,R}=0.305$ for the Cousins $R$ bandpass, 
$\mu_{1,V}=0.412$, and $\mu_{2,V}=0.288$ for the Johnson-Morgan $V$ bandpass,
$\mu_{1,r}=0.344$, and $\mu_{2,r}=0.306$ for the Sloan $r$ bandpass}
based on the values  tabulated in \citet{Claret2011} for the stellar parameters from \citet{Torres2012}.

\begin{figure}
\vspace{0cm}\hspace{0cm}
\includegraphics[width=\columnwidth]{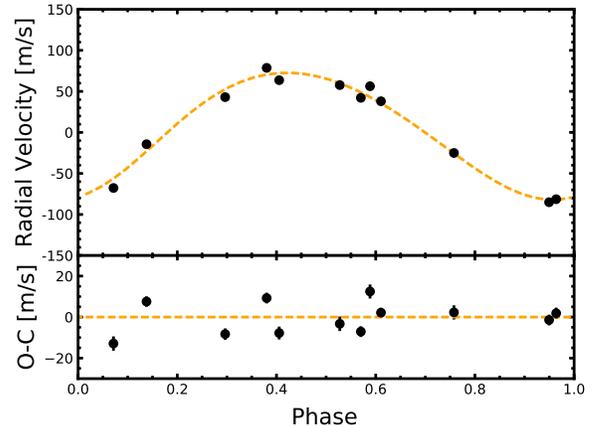}
\caption{Keck/HIRES Doppler velocity measurements of HAT-P-29 from \citet{buchhave2011} and \citet{knutson2014}. 
The best-fitting Keplerian orbit solution from the joint RV and light-curve modelling (see \S3 and Figure~\ref{fig3}) is overplotted as a dashed orange line.
 A constant radial acceleration, $\dot{\gamma}=0.0491_{-0.0086}^{+0.0085}\,\rm{m\,s^{-1}\,day^{-1}}$, is subtracted for clarity.
 The bottom panel shows the residuals of the best fit, which have an RMS scatter of $7.8\,\rm{m\,s^{-1}}$.
\label{fig4}}
\end{figure}
\setlength{\tabcolsep}{1.1pt}

\section{Result and Discussion}

\subsection{Transit Parameters and Physical Properties}

Based on the analysis described above, the physical and orbital parameters for HAT-P-29 system obtained from two global fittings are presented in Table~\ref{table4}.

As suggested by previous studies (e.g. \citealt{Anderson2012, Hartman2015}), the discovery light curves obtained from small telescopes like HATNet \citep{Bakos2004}, WASP \citep{Pollacco2006}, KELT \citep{Pepper2007}, CSTAR \citep{Wang2014} are usually with large PSFs, and often contain contaminating light from nearby stars. In addition, flattening routines such as TFA \citep{Kovacs2005} or SYSREM \citep{Tamuz2005}, which are required to remove red noise from ground-based multi-month datasets \citep{Pont2006}, often affect the observed transit depth as well. Therefore, although two global fitting results agree with each other very well, we consider the one based on only follow-up photometry as our final result (overplotted in Figures~\ref{fig3} and \ref{fig4}). The following discussion is based on this result.

For comparison, the system parameters, estimated in previous studies \citep{buchhave2011, knutson2014}, are also listed in Table~\ref{table4}.

We find almost identical Doppler velocimetric properties to \citet{knutson2014}, as expected, given that the same RVs were used. The results are also in agreement with those from the discovery work \citep{buchhave2011}, and are consistent with zero eccentricity, as one would expect from a tidally circularized hot Jupiter. 

The known RV trend in the HAT-P-29 system, which was previously reported in \citet{knutson2014}, can also been seen in our fitting result. It is believed to be caused by an additional companion with a mass between $1-200\,\mj $ and an orbital separation of $2\,{\rm AU}<a<36\,$AU  \citep{knutson2014}. No further significant RV signal is present in the residuals to our one planet + drift fit, which has a residual RMS scatter of $7.8\,{\rm m\,s^{-1}}$ and which allows us to place constraints on the mass and the period of an additional companion in the system \citep{Wright2007}.

\begin{figure}
\vspace{0cm}\hspace{0cm}
\includegraphics[width=\columnwidth]{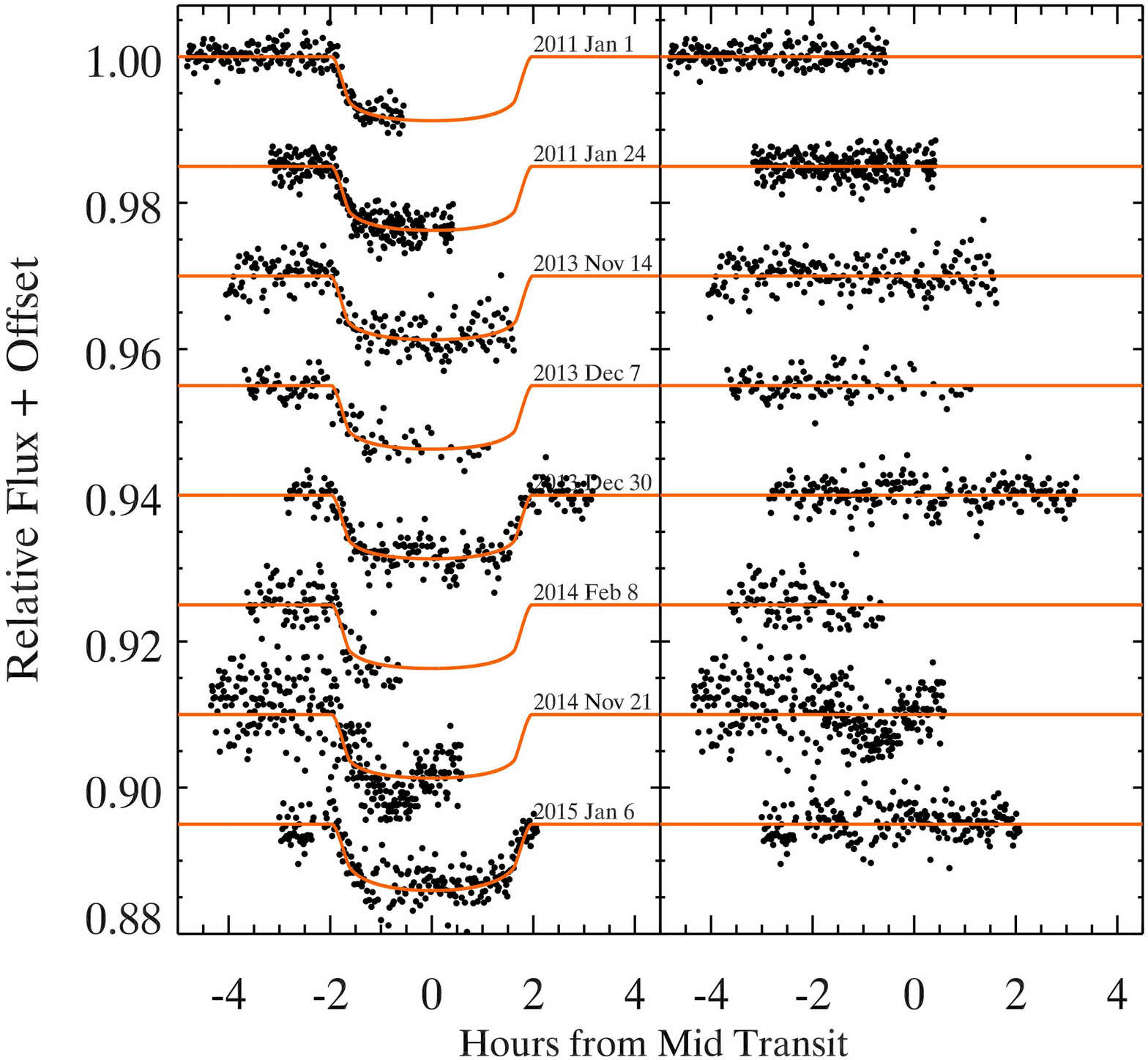}
\caption{
Relative photometry of HAT-P-29 during eight different transits, obtained by \citet{buchhave2011} (top two light curves) and this work (bottom six light curves). The best-fitting model, is shown as a solid orange line.
The residuals appear to the right of each data set. Both light curves and residuals are offset artificially for clarity. See Table~\ref{table2} for further details of each light curve.
\label{fig5}}
\end{figure}

Compared to the transit parameters obtained by \citet{buchhave2011}, we find a slightly different solution, with a smaller planet-to-star radius ratio ($1.4\,\sigma$), a higher orbital inclination ($1.2\,\sigma$), and a correspondingly smaller impact parameter ($1.3\,\sigma$). The orbital inclination is observationally strongly tied to the transit's total duration. Our results, therefore, point to a longer transit duration (by $2.8\,\sigma$).

\begin{figure}
\vspace{0cm}\hspace{0cm}
\includegraphics[width=\columnwidth]{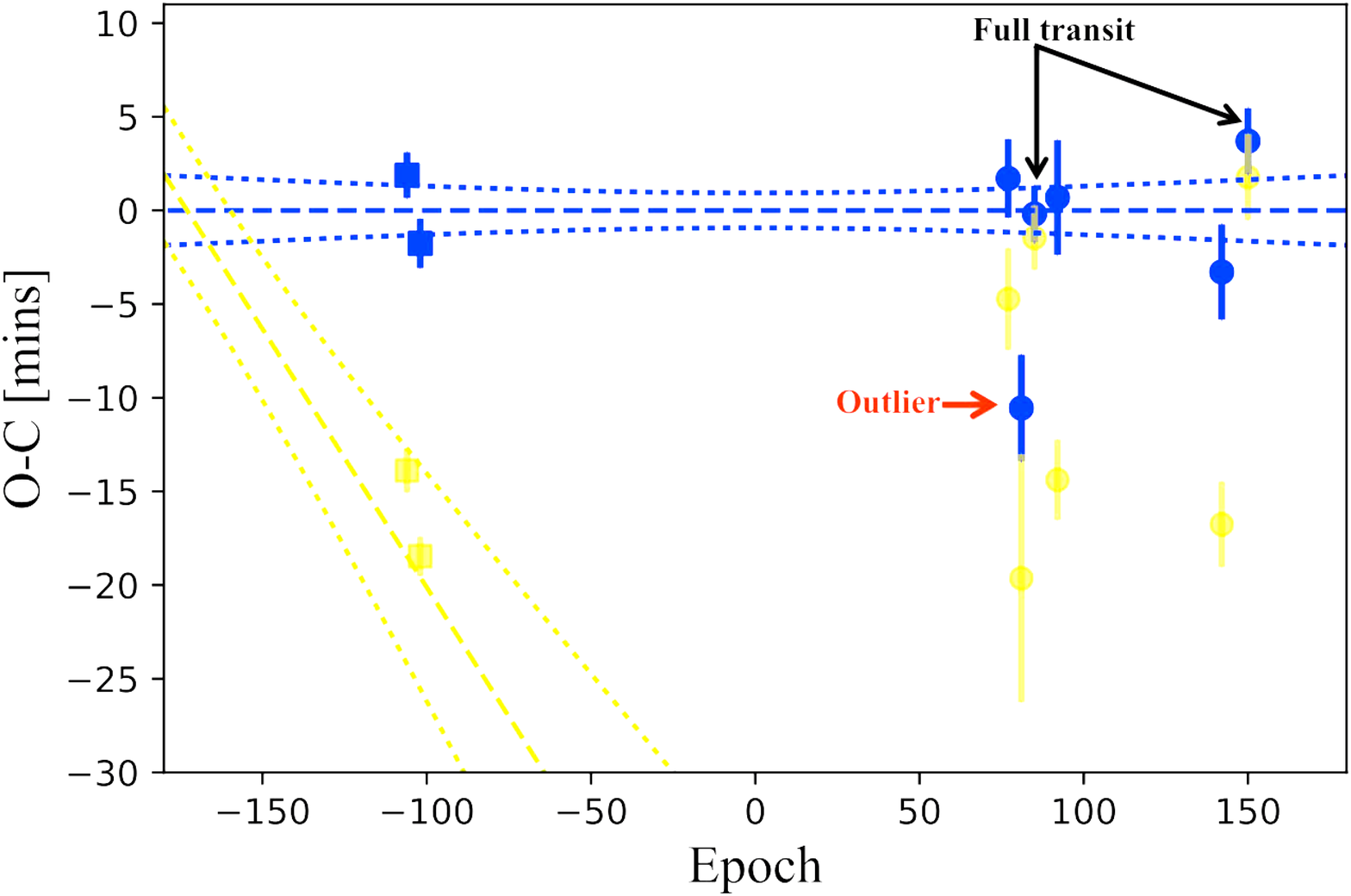}
\caption{
Transit timing residuals for HAT-P-29b, according to the updated linear ephemeris (blue dashed line) given in Table~\ref{table4}.
The transit mid-times derived from our revised transit model are given by the blue markers.
No statistically significant TTVs are detected at a level above $\rm{2.5\,mins}$, except a $4.5\,\sigma$ outlier (indicated by the red arrow) from the UTC 2013 December 7 transit.
The yellow markers indicate the transit mid-times derived using the transit model from \citet{buchhave2011}, which offered an ill-determined transit duration of $202.61\,\rm{mins}$, $30.96\,\rm{mins}$ shorter than that from our fit.
The measured transit mid-times for full transit light curves (indicated by the black arrow) are insensitive to the model duration.
The blue (yellow) dotted lines indicate the propagation of $\pm1\,\sigma$ errors in the linear ephemeris obtained in this (discovery) work. 
\label{fig6}}
\end{figure}

Most importantly, we note that our orbital period is $17.63\,{\rm s}$ longer than the previous measurement, a difference of $4.0\,\sigma$. Our predicted mid-time of the next transit event (on UTC 2018 June 1) is $2.5\,\rm{hrs}$ later than expected according to the orbital ephemeris from \citet{buchhave2011}.

Our new results are based on more extensive photometric data than previous study, so should be more reliable. As shown in \citet{Southworth2012, Benneke2017, WangX2018}, limited aggregations of follow-up photometry are sufficient to confirm the planetary nature, but are rather inaccurate for estimating the system parameters.

To demonstrate that the period discrepancies we found do not arise from differences in the fitting process, we also obtained a global fit that is based only on the data from discovery paper \citep{buchhave2011}. As shown in Table~\ref{table4}, we successfully recover the result from \citet{buchhave2011}, with the two results displaying excellent agreement.


As expected, given the concordant stellar spectroscopic parameters that were employed \citep{buchhave2011, Torres2012}, the stellar properties (stellar mass $M_*$ and stellar radius $R_*$) that emerge from our analysis show good agreement with those of \citet{buchhave2011}.

We use our derived photometric, velocimetric, and stellar parameters to infer the physical properties of HAT-P-29b using the method outlined in \citet{Eastman2013}. The physical properties we find for HAT-P-29b agree with those of \citet{buchhave2011} within uncertainties. 

\subsection{Transit Mid-Times}


To measure the individual transit mid-times, we perform a separate fit to each of six new light curves collected for this work and to the two follow-up light curves obtained in the discovery paper \citep{buchhave2011}. 
We hold all of the global parameters fixed to the value determined from the joint analysis and fit each transit light curve separately, by allowing \textit{only} the transit mid-time ($T_{\rm C}$) as well as the light-curve specific baseline flux ($F_0$) to float.

The resulting transit mid-times ($T_{\rm C}$), measured independently to each new and catalogued light curve, can be found in Table~\ref{table5}.


The RMS deviations of these transit mid-times from our updated linear ephemeris is about $275.2\,{\rm s}$. This value, however, is significantly affected by the $4.5\,\sigma$ outlier from the UTC 2013 December 7 transit which was observed duing poor weather conditions, thereby producing a less precise measurement of the transit mid-time. With this weather-affected outlier removed, no statistically significant TTVs are detected at a level above $148.8\,{\rm s}$.

Figure~\ref{fig6} shows the deviations of transit mid-times (blue markers) for HAT-P-29b from the linear orbital ephemeris determined in this work (blue dashed line) and the discovery work (yellow dashed line). 

Given the period discrepancy discussed in \S4.1, it is not surprising that the transit ephemeris in the discovery paper (yellow dashed line) disagreed with the transit mid-times we obtained from our new light curves (blue circles). It is interesting, however, that the discovery paper's transit ephemeris even significantly disagreed with the transit mid-times we found from their own follow-up light curves (blue squares). 

This situation arises because the transit mid-times are determined using our best-fitting model, which gives a $30.96\,{\rm min}$ longer transit duration than \citet{buchhave2011} measured. The transit mid-times for the partial light curves are very sensitive to the transit duration. Transit mid-times (yellow squares) derived using the transit model from \citet{buchhave2011} are therefore in disagreement with the transit mid-times  determined using our best-fitting model (blue squares), but are consistent with their own transit ephemeris (yellow dashed line).

The measured transit mid-times for full transit light curves, however, are insensitive to the model duration. For the two full transit light curves we obtained, the mid-times derived from the best-fitting transit model reported in \citet{buchhave2011} and this work are in good agreement, and are consistent with our updated orbital ephemeris.

\subsection{Limits on an Additional Perturber}

The lack of statistically significant TTVs, together with the absence of extra signal in the RV residuals around the one-planet + drift model provide a constraint on the dynamical properties of hypothetical close-in perturbers to HAT-P-29b.

Using dynamical simulations with the MERCURY6 planetary orbital integrator \citep{Chambers1999}, we place an upper limit on the mass of the hypothetical perturber as a function of its orbital period, based on our transit timing analysis of HAT-P-29b. The RMS deviation of its measured transit mid-times from the updated linear orbital ephemeris is approximately $148.8\,\rm{s}$.  

For our simulations, the two planets are assumed to be coplanar, and are initially set on circular orbits, a configuration which provides the most conservative estimate of the upper mass limit of the hypothetical perturber, as discussed by \citet{Bean2009}, \citet{Fukui2011}, and \citet{Hoyer2011, Hoyer2012}.

We explore a perturber's orbit with a semi-major axis between $0.0032$ and $0.138\,\rm{AU}$ (i.e. the period ratio of perturber and known planet from 1:3 to 3:1) in steps of $0.001\,\rm{AU}$, which is further reduced to $0.0005\,{\rm AU}$ in the proximity to resonance since the largest planetary TTVs are likely to arise in or near resonance \citep{agol2007, holman2005}. 
For each orbital separation analyzed, the approximation for the upper mass limit of the hypothetical perturber is obtained iteratively by linear interpolation with an initial mass guess of $1\,{\rm M_\oplus}$ and with a convergence tolerance for TTVs of $1\,\rm{s}$.

\setlength{\tabcolsep}{1.1pt}
\begin{longrotatetable}
\begin{deluxetable}{lcccccc}
\tabletypesize{\scriptsize}
\tablewidth{0pt}
\tablecaption{System parameters for HAT-P-29\label{table4}}
\tablehead{\colhead{~~~Parameter} & \colhead{Units} & \colhead{All follow-up LCs}& \colhead{HATnet+all follow-up LCs} & \colhead{HATNet+2011 follow-up LCs} & \colhead{\citealt{buchhave2011}} & \colhead{\citealt{knutson2014}}}
\startdata
\sidehead{Stellar Parameters:}                                                                           
                           ~~~$M_{*}$\dotfill &Mass (\msun)\dotfill  & $1.199_{-0.061}^{+0.063}$                        & $1.198_{-0.063}^{+0.065}$                & $1.177_{-0.059}^{+0.064}$               & $1.207 \pm 0.046$                   & $1.207 \pm 0.046$\tablenotemark{a}        \\
                         ~~~$R_{*}$\dotfill &Radius (\rsun)\dotfill  & $1.237_{-0.071}^{+0.077}$                        & $1.229_{-0.073}^{+0.080}$                & $1.176_{-0.071}^{+0.074}$               & $1.224_{-0.075}^{+0.133}$           & ...                                       \\
                     ~~~$L_{*}$\dotfill &Luminosity (\lsun)\dotfill  & $1.92_{-0.25}^{+0.29}$                           & $1.89_{-0.25}^{+0.30}$                   & $1.70_{-0.23}^{+0.26}$                  & $1.84_{-0.26}^{+0.47}$              & ...                                       \\
                         ~~~$\rho_*$\dotfill &Density (cgs)\dotfill  & $0.89_{-0.13}^{+0.15}$                           & $0.91_{-0.14}^{+0.16}$                   & $1.02_{-0.15}^{+0.18}$                  & ...                                 & ...                                       \\
              ~~~$\log(g_*)$\dotfill &Surface gravity (cgs)\dotfill  & $4.332\pm0.044$                                  & $4.337_{-0.045}^{+0.044}$                & $4.368\pm0.045$                         &  $4.34 \pm 0.06$                    &    $4.34 \pm 0.06$\tablenotemark{b}         \\
              ~~~$\teff$\dotfill &Effective temperature (K)\dotfill  & $6115\pm86$                                      & $6112\pm88$                              & $6085\pm87$                             &  $6087 \pm 88$                      &    $6086 \pm 69$\tablenotemark{b}     \\
                              ~~~$\feh$\dotfill &Metalicity\dotfill  & $0.128_{-0.080}^{+0.079}$                        & $0.128_{-0.080}^{+0.079}$                & $0.132\pm0.080$                         &  $0.21 \pm 0.08$                    &    $0.14 \pm 0.08$\tablenotemark{b}             \\
\sidehead{Planetary Parameters:}                                                                                                                                                                                                                                                                
                               ~~~$e$\dotfill &Eccentricity\dotfill  & $0.075_{-0.027}^{+0.029}$                        & $0.073_{-0.028}^{+0.029}$                & $0.066\pm0.028$                         & $0.095 \pm 0.047$                   & $0.061_{-0.036}^{+0.044}$                 \\
    ~~~$\omega_*$\dotfill &Argument of periastron (degrees)\dotfill  & $203_{-36}^{+29}$                                & $201_{-37}^{+29}$                        & $192_{-38}^{+35}$                      & $169 \pm 30$                        & $211_{-65}^{+39}$                         \\
                              ~~~$P$\dotfill &Period (days)\dotfill  & $5.723390     \pm    0.000013 $            & $5.723376 \pm        0.000021$                 & $5.723178_{-0.00010}^{+0.000099}$       & $5.723186 \pm 0.000049$             & ...                                       \\
                       ~~~$a$\dotfill &Semi-major axis (AU)\dotfill  & $0.0665\pm0.0011$                                & $0.0665\pm0.0012$                        & $0.0661_{-0.0011}^{+0.0012}$            & $0.0667 \pm 0.0008$                 & ...                                       \\
                             ~~~$M_{P}$\dotfill &Mass (\mj)\dotfill  & $0.767_{-0.045}^{+0.047}$                        & $0.767_{-0.045}^{+0.046}$                & $0.761_{-0.044}^{+0.045}$               & $0.778_{-0.040}^{+0.076}$           & $0.773_{-0.051}^{+0.052}$                 \\
                           ~~~$R_{P}$\dotfill &Radius (\rj)\dotfill  & $1.064_{-0.068}^{+0.075}$                        & $1.055_{-0.072}^{+0.079}$                & $1.026_{-0.069}^{+0.073}$               & $1.107_{-0.082}^{+0.136}$           & ...                                       \\
                       ~~~$\rho_{P}$\dotfill &Density (cgs)\dotfill  & $0.79_{-0.14}^{+0.17}$                           & $0.81_{-0.15}^{+0.18}$                   & $0.87_{-0.15}^{+0.19}$                  & $0.71 \pm 0.18$                     & ...                                       \\
                  ~~~$\log(g_{P})$\dotfill &Surface gravity\dotfill  & $3.224_{-0.057}^{+0.055}$                        & $3.232_{-0.059}^{+0.056}$                & $3.252\pm0.057$                         & $ 3.20 \pm 0.07$                    & ...                                       \\
           ~~~$T_{eq}$\dotfill &Equilibrium Temperature (K)\dotfill  & $1271_{-37}^{+39}$                               & $1266_{-39}^{+41}$                       & $1237_{-38}^{+39}$                      & $1260_{-45}^{+64}$                  & ...                                       \\
                       ~~~$\Theta$\dotfill &Safronov Number\dotfill  & $0.0798_{-0.0063}^{+0.0066}$                     & $0.0806_{-0.0066}^{+0.0068}$             & $0.0831_{-0.0066}^{+0.0072}$            & $0.077 \pm 0.007$                   & ...                                       \\
               ~~~$\fave$\dotfill &Incident flux (\fluxcgs)\dotfill  & $0.589_{-0.066}^{+0.076}$                        & $0.581_{-0.068}^{+0.078}$                & $0.529_{-0.063}^{+0.070}$               & $0.569_{-0.075}^{+0.136}$           & ...                                       \\
\sidehead{RV Parameters:}                                                                                                                                                                                                                                                                       
                              ~~~$e\cos\omega_*$\dotfill & \dotfill  & $-0.061_{-0.024}^{+0.025}$                       & $-0.060_{-0.025}^{+0.026}$               & $-0.055_{-0.025}^{+0.027}$              & $-0.084_{-0.046}^{+0.026}$          & $-0.04_{-0.031}^{+0.034}$                 \\
                              ~~~$e\sin\omega_*$\dotfill & \dotfill  & $-0.024_{-0.044}^{+0.037}$                       & $-0.021_{-0.042}^{+0.036}$               & $-0.009_{-0.039}^{+0.035}$              & $0.016 \pm 0.058$                   & $-0.02_{-0.057}^{+0.038}$                 \\
           ~~~$T_{P}$\dotfill &Time of periastron (\bjdtdb)\dotfill  & $2456464.10_{-0.57}^{+0.48}$                     & $2456406.85_{-0.58}^{+0.48}$             & $2455525.29_{-0.60}^{+0.57}$            & ...                                 & ...                                       \\
                    ~~~$K$\dotfill &RV semi-amplitude (m/s)\dotfill  & $77.4_{-3.6}^{+3.8}$                             & $77.5\pm3.6$                             & $77.7\pm3.7$                            & $78.3 \pm 5.9$                      & $77.6_{-4.6}^{+4.5}$                      \\
                 ~~~$M_P\sin i$\dotfill &Minimum mass (\mj)\dotfill  & $0.767_{-0.045}^{+0.047}$                        & $0.767_{-0.045}^{+0.046}$                & $0.760_{-0.044}^{+0.045}$               & ...                                 & ...                                       \\
                       ~~~$M_{P}/M_{*}$\dotfill &Mass ratio\dotfill  & $0.000611_{-0.000030}^{+0.000031}$               & $0.000612\pm0.000030$                    & $0.000617\pm0.000031$                   & ...                                 & ...                                       \\
               ~~~$\gamma$\dotfill &Systemic velocity (m/s)\dotfill  & $10.9\pm4.6$                                     & $10.6\pm4.5$                             & $11.0_{-4.5}^{+4.4}$                    & ...                                 & ...                                       \\
              ~~~$\dot{\gamma}$\dotfill &RV slope (m/s/day)\dotfill  & $0.0491_{-0.0086}^{+0.0085}$                     & $0.0485_{-0.0084}^{+0.0085}$             & $0.0499\pm0.0083$                       & ...                                 & $0.0498_{-0.01}^{+0.0092}$                \\
\sidehead{Primary Transit Parameters:}                                                                                                                                                                                                                                                         
                ~~~$T_C$\dotfill &Time of transit (\bjdtdb)\dotfill  & $ 2456170.5494 \pm 0.0015$                        & $2456445.2740 \pm 0.0022$    & $2455523.7982_{-0.0027}^{+0.0030}$      & $ 2455197.57540 \pm 0.00181 $       & ...                                       \\
~~~$R_{P}/R_{*}$\dotfill &Radius of planet in stellar radii\dotfill  & $0.0885_{-0.0011}^{+0.0012}$                     & $0.0883_{-0.0012}^{+0.0013}$             & $0.0897\pm0.0014$                       & $0.0927 \pm 0.0028$                 & ...                                       \\
     ~~~$a/R_{*}$\dotfill &Semi-major axis in stellar radii\dotfill  & $11.56_{-0.59}^{+0.62}$                          & $11.64_{-0.62}^{+0.63}$                  & $12.09_{-0.63}^{+0.67}$                 & $11.70_{-0.97}^{+0.71}$             & ...                                       \\
              ~~~$u_1$\dotfill &linear limb-darkening coeff\dotfill  & $0.260\pm0.042$                                  & $0.262\pm0.041$                          & $0.259_{-0.047}^{+0.048}$               & 0.2273                              & ...                                       \\
           ~~~$u_2$\dotfill &quadratic limb-darkening coeff\dotfill  & $0.293_{-0.048}^{+0.049}$                        & $0.298\pm0.051$                          & $0.296_{-0.049}^{+0.048}$               & 0.3581                              & ...                                       \\
                      ~~~$i$\dotfill &Inclination (degrees)\dotfill  & $87.99_{-0.56}^{+0.61}$                          & $88.06_{-0.59}^{+0.78}$                  & $87.22_{-0.44}^{+0.47}$                 & $ 87.1_{-0.7}^{+0.5}$               &  $ 87.1_{-0.7}^{+0.5}$\tablenotemark{a}   \\
                           ~~~$b$\dotfill &Impact Parameter\dotfill  & $0.417_{-0.12}^{+0.096}$                         & $0.40_{-0.15}^{+0.10}$                   & $0.594_{-0.088}^{+0.067}$               & $0.591_{-0.094}^{+0.062}$           & ...                                       \\
                         ~~~$\delta$\dotfill &Transit depth\dotfill  & $0.00782_{-0.00020}^{+0.00022}$                  & $0.00779_{-0.00022}^{+0.00023}$          & $0.00804\pm0.00025$                     & ...                                 & ...                                       \\
                ~~~$T_{FWHM}$\dotfill &FWHM duration (days)\dotfill  & $0.1463_{-0.0013}^{+0.0012}$                     & $0.1462_{-0.0013}^{+0.0012}$             & $0.1221_{-0.0071}^{+0.0076}$            & ...                                 & ...                                       \\
          ~~~$\tau$\dotfill &Ingress/egress duration (days)\dotfill  & $0.0157_{-0.0016}^{+0.0021}$                     & $0.0155_{-0.0018}^{+0.0021}$             & $0.0171_{-0.0020}^{+0.0023}$            & $0.0177 \pm 0.0024$                 & ...                                       \\
                 ~~~$T_{14}$\dotfill &Total duration (days)\dotfill  & $0.1622_{-0.0017}^{+0.0018}$                     & $0.1618_{-0.0017}^{+0.0018}$             & $0.1392_{-0.0066}^{+0.0072}$            & $0.1407 \pm 0.0074$                 & ...                                       \\
      ~~~$P_{T}$\dotfill &A priori non-grazing transit prob\dotfill  & $0.0773_{-0.0059}^{+0.0062}$                     & $0.0770_{-0.0057}^{+0.0062}$             & $0.0748_{-0.0055}^{+0.0059}$            & ...                                 & ...                                       \\
                ~~~$P_{T,G}$\dotfill &A priori transit prob\dotfill  & $0.0923_{-0.0071}^{+0.0075}$                     & $0.0920_{-0.0069}^{+0.0075}$             & $0.0895_{-0.0066}^{+0.0071}$            & ...                                 & ...                                       \\
                            ~~~$F_0$\dotfill &Baseline flux\dotfill  & $0.999946\pm0.000073$                            & $0.999944\pm0.000071$                    & $1.000021_{-0.000087}^{+0.000088}$      & ...                                 & ...                                       \\
\sidehead{Secondary Eclipse Parameters:}                                                                                                                                                                                                                                                       
              ~~~$T_{S}$\dotfill &Time of eclipse (\bjdtdb)\dotfill  & $2456465.080_{-0.086}^{+0.091}$                  & $2456407.852_{-0.090}^{+0.094}$          & $2455526.458_{-0.090}^{+0.097}$         & $2455200.132 \pm 0.138$             & ...                                      \\
                       ~~~$b_{S}$\dotfill &Impact parameter\dotfill  & $0.393_{-0.11}^{+0.085}$                         & $0.383_{-0.14}^{+0.091}$                 & $0.576_{-0.085}^{+0.076}$               & ...                                 & ...                                      \\
              ~~~$T_{S,FWHM}$\dotfill &FWHM duration (days)\dotfill  & $0.1410_{-0.0094}^{+0.0087}$                     & $0.1414_{-0.0092}^{+0.0087}$             & $0.1202_{-0.0073}^{+0.0087}$            & ...                                 & ...                                      \\
        ~~~$\tau_S$\dotfill &Ingress/egress duration (days)\dotfill  & $0.0148_{-0.0016}^{+0.0019}$                     & $0.0146_{-0.0016}^{+0.0020}$             & $0.0164_{-0.0021}^{+0.0027}$            &  $ 0.0183 \pm 0.0074$               & ...                                      \\
               ~~~$T_{S,14}$\dotfill &Total duration (days)\dotfill  & $0.1560_{-0.011}^{+0.0098}$                      & $0.1563_{-0.010}^{+0.0095}$              & $0.1371_{-0.0077}^{+0.0088}$            &  $0.1424 \pm 0.0107$                & ...                                      \\
      ~~~$P_{S}$\dotfill &A priori non-grazing eclipse prob\dotfill  & $0.0812_{-0.0037}^{+0.0044}$                     & $0.0805_{-0.0042}^{+0.0045}$             & $0.0765_{-0.0040}^{+0.0042}$            & ...                                 & ...                                      \\
                ~~~$P_{S,G}$\dotfill &A priori eclipse prob\dotfill  & $0.0969_{-0.0046}^{+0.0055}$                     & $0.0961_{-0.0052}^{+0.0056}$             & $0.0916_{-0.0049}^{+0.0051}$            & ...                                 & ...                                      \\
\enddata                                                                                                                                        
\tablenotetext{a}{In \citet{knutson2014}, the host stellar mass ($M_{*}$) and orbital inclination ($i$) for the HAT-P-29 system were adopted from \citet{buchhave2011}.}
\tablenotetext{b}{In \citet{knutson2014}, the stellar spectroscopic properties for HAT-P-29 were adopted from \citet{Torres2012}.}
\end{deluxetable}
\end{longrotatetable}





\setlength{\tabcolsep}{1.1pt}
\begin{deluxetable}{cccc}
\tablewidth{0pt}
\tablecaption{Transit Mid-Times for HAT-P-29b \label{table5}}
\tablehead{
\colhead{Epoch Number} & \colhead{$T_{\rm C}$} & \colhead{$\sigma_{T_{\rm C}}$} & \colhead{$O-C$} \\
\colhead{}   &  \colhead{(${\rm BJD_{TDB}}$)} & \colhead{(second)} &  \colhead{(second)} \\
}
\startdata   
 -106  &  2455563.87156  &   55.89  &   125.23 \\
 -102  &  2455586.76257  &   53.09  &  -95.59 \\
    77  &  2456611.2506    &   120.41    &   9.52   \\
    81  &  2456634.1356    &   253.52    &  -728.18   \\
    85  &  2456657.0363  &   94.53  &  -109.24   \\
    92  &  2456697.1006    &   143.24    &    -60.05    \\
   142 &  2456983.2670    &   368.96    &  -327.55     \\
   150 &  2457029.05882    &   58.88    &   76.44 \\
\enddata
\end{deluxetable}

\begin{figure*}
\vspace{0cm}\hspace{0cm}
\includegraphics[width=\textwidth]{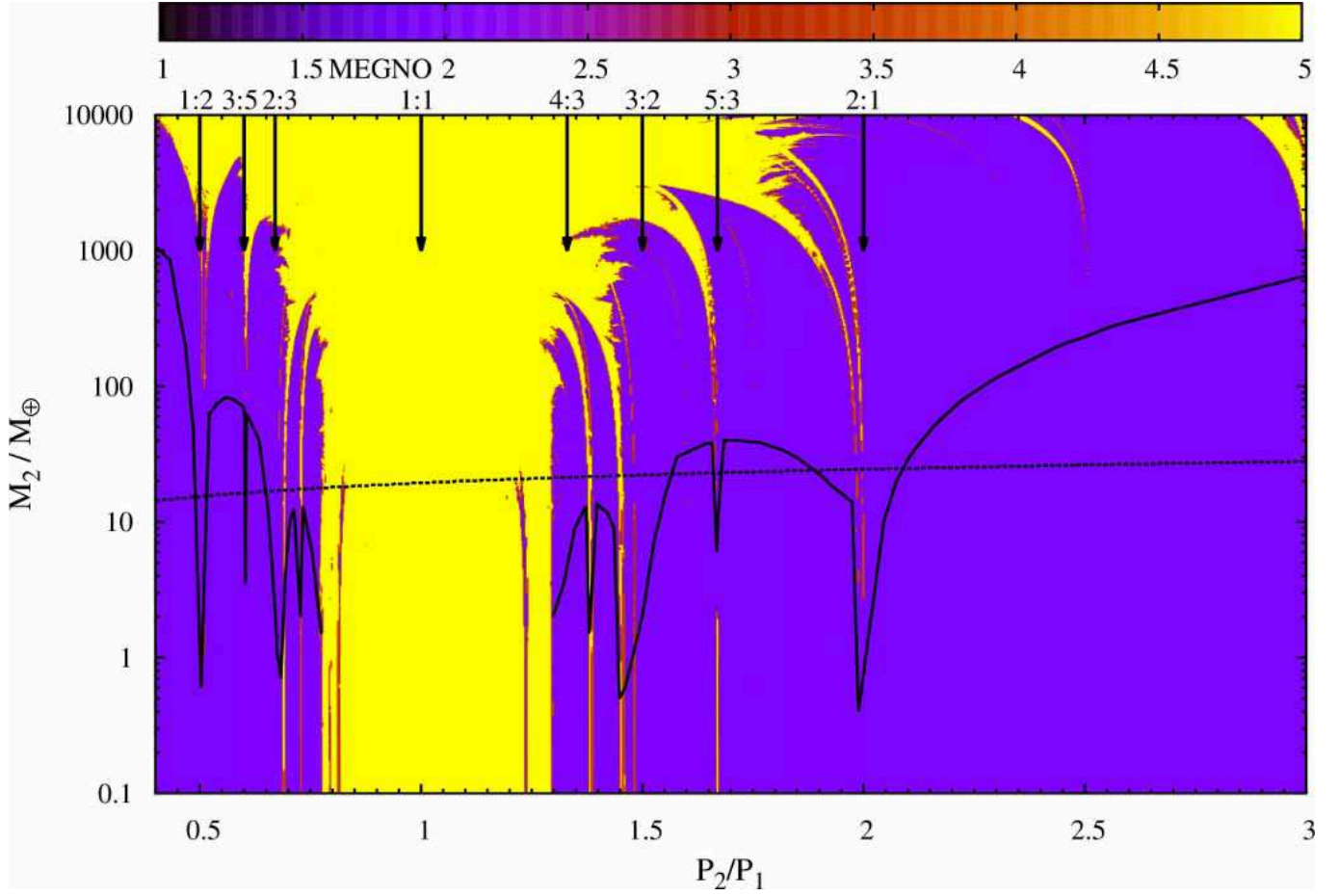}
\caption{Upper mass limit of a hypothetical perturber as a function of the orbital period ratio of the perturber ($P_2$) and the transiting hot Jupiter HAT-P-29b ($P_1$). The solid line corresponds to the mass-period profile where a perturber can produce measured TTVs with an RMS of ~$148.8\,$s. These mass constraints are restrictive near the low-order mean-motion resonances (nominal resonance locations are marked by arrows), where perturbers as small as the Earth are potentially detectable. The dashed line corresponds to the mass limits imposed by the RV measurements. With the RV residuals to the best fit ($\rm{RMS=7.8\,m\,s^{-1}}$), we can exclude the presence of a close-in companion with a mass greater than Neptune. The color-coding delineates the dynamical properties of the three-body system according to the calculated MEGNO factor for a given initial condition. In general, for large MEGNO ($>5$ values with yellow color coding) the system is chaotic. For MEGNO values around 2 (blue color coding) the system's time evolution is quasi-periodic or regular and is usually ascribed to stable motion. It should, however, be pointed out that MEGNO (as with any other numerical chaos indicator) does not provide a proof for orbital stability (quasi-periodic/regular motion). Regular motion is probed only up to the numerical integration time. For longer times the system could, in principle, evolve chaotically. For large values of MEGNO (yellow color coding in the figure with MEGNO $>5$) the system is judged chaotic and is often associated with orbital resonance dynamics (resonance overlap). Chaotic motion often produces unstable orbits, but the system does not necessarily evolve towards an instability as a result of chaos (sticky chaos). Finally, we point out that the locations of orbital resonances between the two planets coincide with the TTV's sensitivity curve. This reiterates that the amplitude of the TTV signal is stronger for systems near orbital resonance.
\label{fig7}}
\end{figure*}

The upper limits on the mass of the hypothetical perturber in the HAT-P-29 system that are determined by these simulations are illustrated in Figure~\ref{fig7}.
While the Doppler residuals with an RMS of $7.8\,\rm{m\,s^{-1}}$ provide stronger constraints on the ``maximum minimum mass'' \citep{Wright2007} of the hypothetical perturber on most configurations (dashed line in the figure), the mass constraints from the TTVs technique (solid black line) are more restrictive at the low-order mean-motion resonances. We can rule out the presence of a perturber with mass greater than $0.6$, $0.7$, $0.5$, and $0.4\,{\rm M_\oplus}$ near the $1:2$, $2:3$, $3:2$, and $2:1$ resonances, respectively.

According to the Mean Exponential Growth of Nearby Orbits  (MEGNO) Index \citep{Gozdziewski2001, Cincotta2003, Hinse2010}, we also show the chaotic/quasi-periodic dynamics for the three-body system in the same figure. The resulting MEGNO map and, in particular the dynamical (chaotic) properties in the vicinity of the transiting planet (large mutual perturbations), is qualitatively consistent with the orbital stability limits derived using the method outlined in \citep{Barnes2006}.

\section{SUMMARY AND CONCLUSIONS}

Planet ``hunting'' is gradually losing its cachet, and is being supplanted by renewed efforts to characterize the known planetary inventory. We, therefore, have initiated a ground-based photometric follow-up project, TEMP, to aid this broader effort, hopefully helping to foster a new understanding of the formation, and evolution of exoplanetary systems.

In our inaugural efforts, we have presented photometry of six transits of HAT-P-29b, obtained between 2013 October and 2015 January with two different telescopes, 
which quadruples the number of published transit light curves available for this planetary system to date. 
The new light curves have photometric scatter ranging from 1.9 to $3.7\,\rm{mmag}$ and a typical exposure time of $35-100\,\rm{s}$.
 
We analyzed our new photometric data, along with two follow-up light curves presented in the discovery paper \citep{buchhave2011}, the RV measurements \citep{buchhave2011, knutson2014}, and improved spectroscopic properties of the host star \citep{Torres2012} to confirm and refine the orbital and physical properties of the HAT-P-29 system. Our improved orbital period is $17.63 \pm 4.38\,$s longer than previous measurements \citep{buchhave2011}, a difference of $ 4.0\,\sigma$, facilitating future characterization of the system during the transit (e.g. the wavelength-dependent transmission spectrum and/or the Rossiter-McLaughlin measurements).

The lack of TTVs with a standard deviation larger than $148.8\,\rm{s}$ placed an upper limit on the mass of a nearby hypothetical perturber as a function of its orbital separation. These mass constraints are particularly restrictive at the low-order mean-motion resonances. Near the $1:2$, $2:3$, $3:2$, and $2:1$ resonances with HAT-P-29b, perturbers with masses greater than $0.6$, $0.7$, $0.5$, and $0.4\,{\rm M_\oplus}$ can be excluded, respectively. Away from mean-motion resonance, the RV residuals, with an RMS of $7.8\,\rm{m\,s^{-1}}$, indicate HAT-P-29 system could readily be harboring additional short-period Neptune-mass companions. Thus, further observations of HAT-P-29, both through photometry and Doppler velocimetry, would be useful in helping to assess the presence of additional nearby planets in the system, especially in dynamically stable non-resonant orbits. The presence or absence of such planets provides direct insight into the formation and evolution processes of hot Jupiters.

For the coming flood of planetary candidates from $K2$ and $TESS$ that require photometric follow-up observations, 
we plan to involve more telescopes in our project to obtain high-precision photometric light curves with the aim to improve physical and orbital properties of transiting exoplanetary systems with poor data coverage. 
In particular we plan to make dramatic progress towards sub-mmag photometric precision by the use of the autoguider and the beam-shaping diffusers \citep{Stefansson2017}.\\

\textbf{Acknowledgments}  \\

We thank the anonymous referee for the insightful suggestions that greatly improved this manuscript.

S.W. thanks the Heising-Simons Fundation for their generous support.

This work is supported by the National Basic Research Program of China (Nos. 2014CB845704, and 2013CB834902);
the National Natural Science Foundation of China (Grant No. 11503009, 11333002, 11673011, 11373033, 11433005, 11673027, 11633009);
the CAS ``Light of West China'' program (2015-XBQN-A-02);
JSPS KAKENHI Grant Number JP18H01265; 
the National Defense Science and Engineering Bureau civil spaceflight advanced research project (D030201);
Minor Planet Foundation of Purple Mountain observatory.

T. C. H. acknowledges KASI research grant 2016-1-832-01.
Numerical computations were partly carried out using the SFI/HEA Irish Center for High-End Computing (ICHEC) and the 3rd generation Polaris High-Performance Computing cluster at KASI/South Korea.

\end{document}